\documentclass[11pt]{article}
\usepackage{pst-all}
\usepackage{pstricks}
\usepackage{pstcol,pst-fill,pst-grad}
\usepackage{amsfonts}
\usepackage{graphicx}
\usepackage{amsmath}
\usepackage{color}

\makeatletter \@addtoreset{equation}{section}

\newcommand{\be}{\begin{equation}}
\newcommand{\ee}{\end{equation}}
\newcommand{\bea}{\begin{eqnarray}}
\newcommand{\eea}{\end{eqnarray}}
\begin{document}
\date{}
\title{
\textbf{       Stringy Dyonic    Solutions    and  Clifford  Structures  }\\
\textbf{   } }
\author{  A. Belfakir$^1$, A. Belhaj$^1$,  Y. El Maadi$^1$, S. E. Ennadifi$^2$,   Y. Hassouni$^1$, A. Segui$^3$
\hspace*{-8pt} \\
{\small  $^1$  Equipe des Sciences de la Mati\`ere et du Rayonnement(ESMAR),  D\'epartement de Physique }\\ {\small Facult\'e des Sciences,  Universit\'e Mohammed V,  Rabat, Morocco}\\
{\small  $^2$   D\'epartement de Physique,   Facult\'e des Sciences,  Universit\'e Mohammed V,  Rabat, Morocco}\\
{\small  $^3$ Departamento de F\'{\i}sica Te\'{o}rica, Universidad de Zaragoza, E-50 009-Zaragoza, Spain} }  \maketitle
\begin{abstract}
Using  the toroidal compactification  of string theory on $n$-dimensional tori, ${\mathbb{T}^n}$,    we  investigate  dyonic  objects in arbitrary dimensions.  First, we  present  a  class of dyonic black solutions formed by two different D-branes using
 a correspondence between toroidal cycles  and  objects  possessing  both magnetic and electric charges, belonging to $ {\mbox{U}}(1)_e^{2^{n-1}}\times {\mbox{U}}(1)_m^{2^{n-1}} $ dyonic  gauge symmetry.   This  symmetry could be  associated with  electrically charged
magnetic monopole solutions  in stringy model buildings of the standard model  extensions.
Then,  we  consider  in some details such black hole classes obtained from  even  dimensional toroidal compactifications, and we  find
that they are linked to  $Cl(n)$ Clifford algebras using the  vee product.
It  is   believed   that this analysis   could  be  extended to  dyonic  objects  which can be obtained from  local Calabi-Yau manifold compactifications.
\end{abstract}
\newpage{}

\thispagestyle{empty}

\newpage{}\setcounter{page}{1} \newpage{}

\section{Introduction}
Recently,  black holes and their extensions  have been extensively investigated  in
connection  with  higher dimensional supergravity  models
compactified on Calabi-Yau (CY) manifolds\cite{1,2,3,4,5}. In certain
compactifications, the scalar field contributions  can be fixed in terms of the black
brane  potential by optimizing  the stringy moduli parameterized by
physical and geometrical deformations. In this way, the
corresponding entropy functions have been  computed   using the
duality symmetries  which act  on the invariant black object
charges. Considering D-brane physics, several
compactfications producing some  black brane solutions embedded in
type II superstrings have been studied   \cite{6,7,8}.  It has been shown    that these black
objects can be linked  to many subjects including  quantum information theory by exploiting the
qubit mathematical formalism \cite{9,10,11,12,13,14,15,16}.  Concretely, a remarkable  correspondence
between  qubit systems  and black holes in superstring
theory has been established. The relevant findings concern a nice
link between   the $N = 2$ STU black hole  with eight charges and
three-qubit system  states \cite{13,17,18,19}. The analysis  has been developed to
describe structures going beyond such  extremal black hole solutions
in terms of higher dimensional qubit systems.  Using string dualities between type
IIA and heterotic superstrings, four-qubit systems in the context of type II
superstring compactifications  have been  investigated \cite{20,21}.  It has been revealed that
such qubit  systems are  related   to a stringy moduli space
$\frac{SO(4,4)}{SO(4)\times SO(4)}$   considered  as  a reduction of
the  moduli space of   six dimensional supergravity model  obtained from the compactification of   the heterotic superstring on  $\mathbb{T}^4$ \cite{22}. Moreover,  the
string/string duality  has been exploited to show that  there exists an interplay between   four-qubit
states  and  a  particular ordinary  dyonic black  hole in six dimensions having eight electric and eight  magnetic charges.

 More recently, electric and magnetic charges have been reconsidered   to discuss  a  dark  matter sector  using  the $SL(2, \mathbb{Z})$
duality \cite{220}. It is recalled that  such a sector  involves  a massive dark
photon and dark magnetic monopoles  investigated in  terms of a  kinetic mixing of the dark and
usual  photons. These activities could be related  to  black objects  in the presence
of Dark Energy   being  the hardest puzzles in
modern and  new  physics.   In this regard, any serious  attempt  corresponding to such an ambiguous   dark sector  is
welcome.

The aim of this work is to contribute to this program  by
 reconsidering the study of  dyonic black objects in arbitrary dimensions obtained from the  toroidal compactification  of string theory on ${\mathbb{T}^n}$.  Precisely, we  first  present  a  class of dyonic black solutions, given by doublets,  formed by two different D-branes using
 a correspondence between toroidal cycles  and  objects  carrying electric and magnetic  charges. These charges belong to  $ {\mbox{U}}(1)_e^{2^{n-1}}\times {\mbox{U}}(1)_m^{2^{n-1}} $ dyonic  gauge symmetry.    This  symmetry could be  associated with  electrically charged
magnetic monopole solutions  in stringy model buildings of the standard model (SM) extensions.
 Then, we  pay  a particular attention to  black  object  classes obtained from lower even  dimensional toroidal compactifications. Notably,  we  find
that they are linked to  $Cl(n)$ Clifford algebra structures  using  the vee product of  real differential forms on  ${\mathbb{T}^n}$.
It  is   suggested   that this analysis   could  be  extended to  dyonic  objects obtained from  local Calabi-Yau manifold compactifications.\\
The paper is organized as follows. After a study of  stringy dyonic  black objects in section 2, we
build   the associated 
toroidal compactifications   in section 3.    Then,  we  emphasize the situation for    even dimensional compactifications. A link with dark sectors is provided.  Section
4 concerns a  relation with $Cl(n)$ Clifford algebras using  the vee ($\vee$)  product associated with  real differential forms dual to cycles on which D-branes can  be wrapped on  to generate dyonic  black objects. Concluding remarks and discussions   are presented in section 5.

\section{ Stringy dyonic  black objects in arbitrary dimensions}
In this section, we reconsider the study of     dyonic solutions in
type II superstrings compactified on  $n$-dimensional  real
manifolds $X^n$. It has been remarked   that each  compact manifold possesses    some geometric information    which  can play a
crucial role in the determination of the superstring theory
spectrum in $10-n$ dimensions \cite{22}.  In particular,  they  can
be exploited  to produce  all physical data in lower dimensional
superstring models. In connections with extremal black  objects, it
has been suggested  that the black $p$-branes can be built using  a
system involving  $(p+k)$-branes wrapping  appropriate  $k$-cycles of
$X^n$.  It has been remarked   that the corresponding near horizon geometries  are given by a
product of AdS spaces and spheres $
 Ads_{p+2}\times S^{8-n-p}$ where    $n$ and $p$  are integers   constrained by
$1 \leq n$ and $ 2 \leq 8-n-p$.
It is noted  that these black  objects are  coupled to $(p+1)$-gauge
field $C_{p+1}$. To measure  their electric charges, one should use
the field strength  $dC_{p+1}$. This field  will  play the same role
of  the  field strength $F=dA$, where $A$ is the 1-form  gauge potential coupled to the
ordinary particle associated with the following action term
\begin{equation}
S\sim \int F \wedge \star F,
\end{equation}
where $\star F$ is the dual of $F$. It is known that the electric
charge is measured  by the integration
\begin{equation}
Q_e=\int_{\mathbb{S}^{10-n-2}}\star F,
\end{equation}
where $\mathbb{S}^{10-n-2}$ is   the $(10-n-2)$-dimensional real
sphere. Similarly, one can compute  the magnetic charge  using the relation
\begin{equation}
P_m=\int_{S^{2}}F.
\end{equation}
These  particle  charge equations  can be generalized to $p$-branes
being $p$-dimensional objects moving in the  string theory space-time. In
this case, the electric charge is calculated  as follows
\begin{equation}
Q_e=\int_{\mathbb{S}^{10-n-p-2}}\star dC_{p+1}
\end{equation}
 where $\star dC_{p+1}$ is the dual of  the  $ dC_{p+1}$ gauge field. However, the  magnetic charge can be given by the following integration
\begin{equation}
P_m=\int_{\mathbb{S}^{p+2}}dC_{p+1},
\end{equation}
where $\mathbb{S}^{p+2}$ is   the $(p+2)$-dimensional real
sphere. In higher dimensional theories including string theory, one may classify the black brane
solutions using  the extended electric/magnetic duality connecting a
$p$-dimensional electrical black   brane to a $q$-dimensional
magnetic one  via the following relation
\begin{equation}
\label{duality} p+q=6-n.
\end{equation}
This   constraint  can be solved in different manners  using
 appropriate  $(p,q)$ couple values including the ones representing
dyonic solutions in arbitrary  dimensions.  However, a deeper  inspection  in string compactifications shows  that dyonic solutions carrying  electric and magnetic charges can be  arranged  as  object like doublets
\begin{equation}\left(\begin{array}{c}
p\\
q
\end{array}\right).
\end{equation}
 This doublet configuration is motivated and supported by the $SL(2,Z)$  electric-magnetic duality \cite{220}.  It has been remarked that
these dyonic  solutions  are associated with a  gauge field  symmetry involving two factors corresponding to electric and magnetic sectors
\begin{equation}
\mbox{G}_{dyon}= \mbox{G}_e \times  \mbox{G}_m.
\label{twofactors}
\end{equation}
In the compactification  on $n$-dimensional real manifolds, two different dyonic solutions could appear with electric and magnetic charges.  They are classified in what follows.
\subsection{Ordinary dyonic solution}
 This   solution is described  by the constraint
\begin{equation}
p=
q.
\end{equation}
It turns out that such a solution  finds a place only in even dimensional string theory compactifications.  A simple calculation shows that  one has

\begin{equation}
p=
q= 3-\frac{n}{2}.
\end{equation}
This generates dyons   represented by  the following doublets
\begin{equation}
\left(\begin{array}{c}
p\\
q
\end{array}\right)
=\left(\begin{array}{c}
3-\frac{n}{2}\\
3-\frac{n}{2}
\end{array}\right)
\end{equation} consisting   of the same object which  appears  in even  dimensional superstring models. This solution  can be described only  by a single factor  related to both electric and magnetic charges. In this way, the above  dyonic symmetry can be reduced to
\begin{equation}
\mbox{G}_{dyon}= \mbox{G}_e.
\end{equation}
\subsection{Non ordinary dyonic solution}
The second solution  concerns the case
\begin{equation}
p\neq
q
\end{equation}
associated with two different D-branes. This involves  dyonic objects
 consisting of an electrically charged black object and its magnetic
dual one. In such a case,  $p$ and $q$   are  constrained by
\begin{equation}
 q=6-n-p,\qquad p\neq3-\frac{n}{2}.
\end{equation}
This  solution generates  doublets of the following form   \begin{equation}
\left(\begin{array}{c}
p\\
q
\end{array}\right)
=\left(\begin{array}{c}
p\\
6-n-p
\end{array}\right)
\end{equation}
carrying electric and magnetic charges associated with two  gauge symmetry factors given in (\ref{twofactors}). This new  dyonic solution could be worked out   to provide a possible support of monopoles  in the context of string theory compactifications.  \\
We wish to add some comments on these dyonic  black  solutions. The first
comment concerns the fact that  the non ordinary  solution is considered as a
single object sharing similar features of the ordinary one
described by the same object, appearing in  even dimensional string
theory.  The  associated  dyonic physics is invariant under the
mapping
    \begin{equation}
p \to 6-n-p
\end{equation}
interpreted  as an electric-magnetic symmetry in string theory compactifications. The second comment concerns the brane configurations of such  dyonic
black  solutions.  In string theory,  they should be
    constructed   from  the  following dyonic  brane doublets living in ten dimensional space time
\begin{equation}\left(\begin{array}{c}
D0\\
D6
\end{array}\right), \quad \left(\begin{array}{c}
NS1\\
NS5
\end{array}\right), \quad
\left(\begin{array}{c}
D1\\
D5
\end{array}\right), \quad
\left(\begin{array}{c}
D2\\
D4
\end{array}\right), \quad
\left(\begin{array}{c}
D3\\
D3
\end{array}\right).
\end{equation}
    Before examining a
particular supergravity theory solution, we should recall that the
black object charges depend on the choice of the internal space. It
turns out that  each compactification involves a Hodge diagram
carrying  not only geometric information but also physical data   providing
the   above mentioned dyonic  black  solutions. \section{Dyonic  black solutions from
toroidal compactifications}
 For
simplicity reasons, we consider  the  toroidal compactification
having a  nice real Hodge diagram playing a primordial   role in the
elaboration of dyonic  stringy   solutions   in $10-n$ dimensions. It is recalled
that $\mathbb{T}^n$ is a flat compact space which can be constructed
using different approaches. One of them is to exploit the trivial
circle fibrations given by the identifications $
x_i\equiv x_i+1, \;\; i=1,\ldots,n. $
It is   remarked that the general  real Hodge  diagram  of
$\mathbb{T}^n$ can  encode all possible non trivial cycles
describing geometric data including the  size and the shape
parameters.   Using a binary number notation,  the number
$h^{e_1,\ldots,e_n}$ will be  associated with a real differential form  of
degree $k$ ($k$-differential form)
$\prod_{\ell=1}^{n}(\overline{{e_\ell}}+e_\ell
 dx_\ell)$, where   $e_\ell$ takes  either $0$ or $1$, and where  $\overline{{e_\ell}}$ is  its
 conjugate. In this way, $k$ equals to  $\sum _{i=1}^ne_i$.
To illustrate  such  real  Hodge diagrams, one may consider
lower dimensional cases corresponding to   $n=1$, $n=2$ and $n=3$.
They  are given by, respectively
\[
\begin{tabular}{|l|l|l|}
\hline
$n=1$ & $%
\begin{tabular}{lll}
& $h^{1}$ &  \\
 & $h^{0}$ &
\end{tabular}%
$ & $%
\begin{tabular}{lll}
& $1$ &  \\
 & $1$ &
\end{tabular}%
$ \\ \hline
$n=2$ & $%
\begin{tabular}{lll}
& $h^{1,1}$ &  \\
$h^{1,0}$ &  & $h^{0,1}$ \\
& $h^{0,0}$ &
\end{tabular}%
$ & $%
\begin{tabular}{lll}
& $1$ &  \\
$1$ &  & $1$ \\
& $1$ &
\end{tabular}%
$ \\ \hline
$n=3$ & $%
\begin{tabular}{lllll}
&  & $h^{1,1,1}$ &  &  \\
& $h^{1,1,0}$ & $h^{1,0,1}$ & $h^{0,1,1}$ &  \\
&$h^{1,0,0}$ & $h^{0,1,0}$   & $h^{0,0,1}$ \\
 &  & $h^{0,0,0}$ &  &
\end{tabular}%
$ & $%
\begin{tabular}{lllll}
&  & $1$ &  &  \\
& $1$ & $1$ & $1$ &  \\
&$1$ & $1$   & $1$ \\
 &  & $1$ &  &
\end{tabular}
$ \\ \hline
\end{tabular}%
\]
\def\m#1{\makebox[10pt]{$#1$}}
\subsection{Dyonic black objects on $\mathbb{S}^1$}
To  establish  a link with string theory,  we  consider some dyonic
black objects  obtained from the compactification  on
1-dimensional circle $\mathbb{S}^1$. Indeed, the nine dimensional
dyonic black objects involve  one electric charge $ Q_0$  and one magnetic
charge $  P_0$  associated with   the    dyonic gauge symmetry
\begin{eqnarray}
 \mbox{G}_{dyon}= {\mbox{U}}(1)_e\times {\mbox{U}}(1)_m.
\end{eqnarray}
The  corresponding brane
representations can be obtained from the ten dimensional dyonic
ones illustrated in the previous section. Using the  Hodge diagram representation, the  brane configurations are constrained by
\begin{equation}
p+q= 5.
\end{equation}
In this case, the  dyonic solutions can be classified as
\begin{equation}
\left(\begin{array}{c}
D0\\
D5
\end{array}\right), \quad \left(\begin{array}{c}
D1\\
D4
\end{array}\right), \qquad \left(\begin{array}{c}
D2\\
D3
\end{array}\right).
\end{equation}
These objects  can be built from the following wrapped D-branes, respectively
\begin{equation}
 \begin{tabular}{lll}
& D0 &  \\
 & $\mbox{D}6/\mathbb{S}^1_1$ &
\end{tabular} \quad  \begin{tabular}{lll}
& D1 &  \\
 & $\mbox{D}5/\mathbb{S}^1_1$ &
\end{tabular} \quad \begin{tabular}{lll}
& D2  \\
 &$ \mbox{D}4/\mathbb{S}^1_1.$ &
\end{tabular}
\end{equation}
The  charges of these  solutions  are associated with  the cycles  dual to the
following real forms on the circle
\begin{equation}
1,\quad dx.
\end{equation}
\subsection{Dyonic black objects on $\mathbb{T}^2$} Eight dimensional dyonic
black solutions
can be  obtained from the compactification on  $\mathbb{T}^2$ constrained by  \begin{equation}
p+q= 4.
\end{equation}
In this case,
the corresponding type II superstrings involve three dyonic black
solutions given  by
\begin{equation}\left(\begin{array}{c}
0\\
4
\end{array}\right), \quad \left(\begin{array}{c}
1\\
3
\end{array}\right), \quad\left(\begin{array}{c}
2\\
2
\end{array}\right).
\end{equation}
For instance, the $\left(\begin{array}{c}
0\\
4
\end{array}\right)$  dyonic solution can be represented by the following  D-brane
charge configurations
\begin{equation}
\begin{tabular}{lll}
& \mbox{D}0 &  \\
\mbox{D}1/$\mathbb{S}^1_1$ &  & \mbox{D}5/$\mathbb{S}^1_2$ \\
& \mbox{D}6/$\mathbb{T}^2.$ &
\end{tabular}
\end{equation}
 Exploiting string theory links including the
S-duality, an equivalent D-brane configuration can be obtained by the
 following  mapping
\begin{eqnarray}
\mbox{D}1 &\leftrightarrow \mbox{NS}1 \nonumber\\
\mbox{D}5 &\leftrightarrow \mbox{NS}5.
\end{eqnarray}
However, the remaining  $\left(\begin{array}{c}
p\\
q
\end{array}\right)$  dyonic  solutions  such that $p+q=4$  can be represented by the following  D-brane
charge configurations
\begin{equation}
\begin{tabular}{lll}
& D$p$ &  \\
D$p+1/\mathbb{S}^1_1$ &  & D$q+1/\mathbb{S}^1_2$ \\
& D$q+2/\mathbb{T}^2$ &
\end{tabular}
\end{equation}
having  two electric charges and two magnetic charges. These charges  associated
with  the cycles  belonging to $\mathbb{T}^2=\mathbb{S}^1_1\times
\mathbb{S}^1_2$ dual  to the following real  differential  forms
 on
$\mathbb{T}^2$  can be organized  as follows
\begin{equation}
\begin{tabular}{lll}
& 1 &  \\
$dx_{1}$ &  & $dx_{2}$ \\
& $dx_{1}dx_{2}.$ &
\end{tabular}
\end{equation}
In this representation, the forms (1, $dx_{1}$) correspond to
the (D$p$,  D$p+1/\mathbb{S}^1_1$)  brane system  involving two
electric charges  associated with $ \mbox{G}_{e}= {\mbox{U}}(1)_e^2$. However, the  dual forms ($dx_{2}$,
$dx_{1}dx_{2}$)  are linked to   the (D$q+1/\mathbb{S}^1_2$,
D$q+2/\mathbb{T}^2$)  brane configuration  having two magnetic
charges corresponding to $ \mbox{G}_{m}= {\mbox{U}}(1)_m^2$.  For $p=q=2$,  it  is worth noting that  the dyonic gauge symmetry    $ \mbox{G}_{dyon}= {\mbox{U}}(1)_e^2\times {\mbox{U}}(1)_m^2 $ reduces to  $ \mbox{G}_{dyon}= {\mbox{U}}(1)^2$   for both electric and magnetic charges.
\subsection{Dyonic black objects on $\mathbb{T}^3$} In this
subsection,   we consider  three copies of $\mathbb{S}^1$
corresponding to the compactification of  superstring theory on
$\mathbb{T}^3$. The manifold is  determined by three circles
$\mathbb{S}^1_1$, $\mathbb{S}^1_2$ and $\mathbb{S}^1_3$ coordinated
by $x_1$, $x_2$ and $x_3$, respectively. This compactification  produces seven
dimensional dyonic black solutions having  four electric charges and
four
 magnetic ones, under $ \mbox{G}_{dyon}=  {\mbox{U}}^4(1)_e\times {\mbox{U}}^4(1)_m $  dyonic gauge symmetry. In this
scenario, one has two fundamental solutions given by the following
doublets
\begin{equation}\left(\begin{array}{c}
0\\
3
\end{array}\right), \quad \left(\begin{array}{c}
1\\
2
\end{array}\right).
\end{equation}
Using the  T-duality  and the  real Hodge diagram of $\mathbb{T}^3$, the
solution  $\left(\begin{array}{c}
0\\
3
\end{array}\right)$
can be built using the following  D-brane configuration
\begin{equation}
\begin{tabular}{lll}
& D0 &  \\ \\
D1/$\mathbb{S}^1_1$ & D1/$\mathbb{S}^1_2$ & D1/$\mathbb{S}^1_3$ \\
\\
D5/$\star \mathbb{S}^1_1$ & D5/$\star \mathbb{S}^1_2$ & D5/ $\star
\mathbb{S}^1_3$
\\ \\ & D6/$\mathbb{T}^3$ &
\end{tabular}
\end{equation}
where $\star \mathbb{S}^1_i$ are dual to $ \mathbb{S}^1_i$ in
$\mathbb{T}^3$.
However,  the solution $\left(\begin{array}{c}
1\\
2
\end{array}\right)$ involves only one Hodge  D-brane configuration
given by
\begin{equation}
\begin{tabular}{lll}
& D1 &  \\ \\
D2/$\mathbb{S}^1_1$ & D2/$\mathbb{S}^1_2$ & D2/$\mathbb{S}^1_3$ \\
\\
D4/$\star \mathbb{S}^1_1$ & D4/$\star \mathbb{S}^1_2$ & D4/ $\star
\mathbb{S}^1_3$
\\ \\& $D5/\mathbb{T}^3.$ &
\end{tabular}
\end{equation}
The   four states   $\{D1,
D2/\mathbb{S}^1_1,   D2/\mathbb{S}^1_2,  D2/\mathbb{S}^1_3\}$  represent  seven dimensional dyonic solutions
carrying  electric charges. The remaining four  states  having
   magnetic charges are obtained using the Hodge duality which  has
been used to build the real Hodge diagram illustrated in
section 2. An examination reveals that these  states can be linked
to   the following real
Hodge diagram of $\mathbb{T}^3$
\begin{equation}
 \begin{tabular}{lll}
&$1$ &  \\\\
$dx_1$ & $dx_2$ & $dx_3$ \\
\\
$\star dx_1$ & $\star dx_2 $ & $\star dx_3$ \\  \\ & $ dx_1dx_2dx_3$
&
\end{tabular}
\end{equation}
where $\star dx_i$ are Hodge duals to $dx_i$ in  $\mathbb{T}^3$.

\subsection{Dyonic black objects  on $\mathbb{T}^n$ and dark sectors}
The compactification of superstring  theory
on $\mathbb{T}^n$ generates  dyonic black solutions in $10-n$
dimensions. In this way,  the corresponding real Hodge diagram can be built
in terms of the forms
\begin{equation}
1,\;\;dx_1,\;\;dx_2,\ldots,\;\;dx_1\ldots dx_n.
\end{equation}
Concretely, the   dyonic black object states can be associated with
 such a set of differential forms  dual to cycles in which D-branes can be  wrapped on.  The presence of  dyonic black objects  $\left(\begin{array}{c}
p\\
6-n-p
\end{array}\right)$,  in such a superstring theory compactification,  suggests that   there are  $2^{n-1}$  states  associated
   with the electric charges. By using the Hodge  mapping between forms
\begin{equation}
k-forms \Leftrightarrow  (n-k)-forms,
\end{equation}
we can show that there are also  $2^{n-1}$  states corresponding  to the magnetic  charges assured by the decomposition
   \begin{equation}
2^{n}=2^{n-1}+2^{n-1}
\end{equation}
supported by   the  dyonic symmetry \begin{equation}\mbox{G}_{dyon}={\mbox{U}}(1)_e^{2^{n-1}}\times {%
\mbox{U}}(1)_m^{2^{n-1}}. \end{equation}   This symmetry, associated with the electrically charged
magnetic monopole solutions within the context of superstring theory,   could be
relevant in stringy model buildings of the standard model (SM) extensions \cite{231,232}.  Precisely,
interesting extensions of SM can be constructed from compactified
superstring models where the geometry and the topology of the internal space
give rise to extra abelian gauge symmetries ${\mbox{U}}_{X_{i}}(1)'$s corresponding to  some conserved charges $X_{i}$ along with new scalar fields  $S_{i}$. This  can be  usually achieved in the context of intersecting D-branes wrapping
non trivial cycles embedded in compact manifolds \cite{233,234,235}. Thus, it could
be shown that there are many roads to handle the physics underlying such   a
dyonic  symmetry  $\mbox{G}_{dyon}$  in the stringy inspired extended SM's. For the charge
identification $X_{i}\equiv {Q}_{e},P_{m}$, we can consider the assumed
symmetry such as
\begin{equation}
{\mbox{U}}_{X_{i}}(1)^{2^n}\equiv {\mbox{U}}(1)_e^{2^{n-1}}\times {\mbox{U}}(1)_m^{2^{n-1}},
\end{equation}%
where the new scalar fields  $S_{i}$ can   play an important role in the underlying scale
as well as the resulting mass spectrum. On the basis of such motivations, one can deal with the
extended gauge symmetry
\begin{equation}
{\mbox{G}}_{SM+dyon}={\mbox{SU}}_{C}(3)\times {\mbox{SU}}_{L}(2)\times \mbox{G}_{dyon}
\end{equation}%
where the first piece ${\mbox{SU}}_{C}(3)$ refers to the strong interaction
associated with the color charge $C$. The second one  ${\mbox{SU}}_{L}(2)$ refers
to the weak interaction corresponding to  the left isospin charge $L$. However,
 the sector ${\mbox{SU}}_{L}(2)\times {\mbox{G}}_{Dyon}$ refers now to the extended
electroweak   dark sector. Roughly, this extended electroweak symmetry is expected
to be broken  by the vacuum expectation value (vev) of a new
scalar $S$ down to the four dimensional SM one. This,  in turn,  is broken down
to the electromagnetic symmetry ${\mbox{U}}_{Qem}(1)$ by the vev of the standard
Higgs scalar field $H$ in the following scheme
\begin{equation}
{\mbox{SU}}_{L}(2)\times {{\mbox{U}}}(1)_e^{2^{n-1}}\times {{\mbox{U}}}(1)_m^{2^{n-1}}\overset{%
\left\langle S\right\rangle }{\rightarrow }{\mbox{SU}}(2)_{L}\times {U}(1)_{Y}\overset%
{\left\langle H\right\rangle }{\rightarrow }{{\mbox{U}}}_{Q_{em}}(1),
\end{equation}%
where now the new complex scalar $S$,  associated with the dyonic symmetry breaking,
corresponds to a $2^{n-1}$-tuple with $2^{n}$ real components as $S=
s^1_i+is^{2}_i, \; i=1, \ldots, 2^{n-1} $. At this point, seen that the dyonic symmetry breaking scale is
expected to be above the SM electroweak scale $\left\langle
S\right\rangle  >\left\langle H\right\rangle \sim 10^{2}GeV$,   this  would make such  dyons heavy as being proportional somehow to the new scalar
vev. Therefore, they  are far to be observed at the current experiments unless
the corresponding proportionality is  highly suppressing. In this case, the
best way of probing the existence of such  dyons remains their observation in
cosmic rays as recently inspected by the actual telescopes  \cite{236,237,238,239}.

Having discussed how   a possible link with extended SM is provided by using  extra symmetries and scalar fields.  It has been shown that  these symmetries can be considered as structures of  division
algebraic ladder operators. In what follows, we elaborate a  relation  with  Clifford algebraic structures reported in \cite{26,27}.

\section{Link with  differential form structures}

A close inspection shows that  the dyonic black objects  obtained from  toroidal compactifications   can be linked to  non trivial structures corresponding to  the real  differential forms on  $\mathbb{T}^n$.

To do so, let us first recall such a structure \cite{26,27,24,25,28}. Let  $V$ be a  vector space over the field of real numbers.  In this space, we can define  a   quadratic form  using  the map $Q: V\rightarrow R$, where $Q$ satisfies
\begin{equation}
Q(\alpha v)=\alpha^{2}Q(v)
\end{equation}
for all $\alpha$ in $R$ and $v$ in $V$.  One can also  define the Grassmann-Cartan exterior product $\wedge$ of differential forms as
\begin{equation}
dx^{\mu}\wedge dx^{\nu}=-dx^{\nu}\wedge dx^{\mu},\quad \mu\ne\nu,  \qquad
dx^{\mu}\wedge dx^{\mu}=0.
\end{equation}
The Grassman product is an associative product $(dx^{\mu}\wedge dx^{\nu})\wedge dx^{\lambda}=dx^{\mu}\wedge (dx^{\nu}\wedge dx^{\lambda})$. However,
the $\vee$ product of two diffrential forms is defined as
\begin{equation}
dx^{\mu}\vee dx^{\nu}=(dx^{\mu},dx^{\nu})+dx^{\mu}\wedge dx^{\nu}.
\end{equation}
Here,  $(dx^{\mu},dx^{\nu})$ denotes the scalar product between $dx^{\mu}$ and $dx^{\nu}$ and $(dx^{\mu},dx^{\nu})=g^{\mu\nu}$ where $g^{\mu\nu}$ is the vector space metric. In what follows,  we will take $g^{\mu\nu}=\delta^{\mu\nu}$  by considering  only  the euclidien metrics associated with the toroidal compactification using the vee product. The Clifford algebra $Cl(n)$ generated by the differential forms is defined as
\begin{equation}
dx^{\mu}\vee dx^{\nu}+dx^{\nu}\vee dx^{\mu}=2\delta^{\mu\nu}Q(dx^{\mu})
\end{equation}
where $Q(dx^{\mu})=\lbrace dx^{\mu},dx^{\mu}\rbrace=1$, $\mu=1,\dots, \frac{n}{2}$ and  $n$ is an even number. Using the  string theory  toroidal compactifaction,  we will see that
one can  construct a set of operators $a_{\ell}$ satisfying
\begin{equation}
\lbrace a_{\ell},a_{\ell'}\rbrace=\lbrace a_{\ell}^{\dagger},a_{\ell'}^{\dagger}\rbrace=0,  \quad \lbrace a_{\ell},a_{\ell'}^{\dagger}\rbrace=\delta_{\ell\ell'}.
\end{equation}
\subsection{$Cl(n)$ and differential  forms on $\mathbb{T}^n$ }
To make contact with the above structure, we will be interested in the  complex geometry associated with $n$ even.  For $n=2$,  the  operators $a$ and $a^{\dagger}$
are given   in  terms of one-forms on  $\mathbb{T}^2$.  Indeed, one has the following identifications
\begin{equation}
a=\frac{1}{2}(dx^{1}+idx^{2}), \quad a^{\dagger}=\frac{1}{2}(-dx^{1}+idx^{2}).
\end{equation}
It is remarked that $\dagger$ maps $i\longrightarrow-i$ and $dx^{j}\longrightarrow-dx^{j}$ for $j=(1,2)$.
The $Cl(4)$ Clifford  algebra  could be built using  the  real differential form set  $\lbrace dx^{1},dx^{2}, dx^{3},dx^{4}\rbrace$. From these forms,  we construct a set of operators satisfying  the above structure. In this case,  we  have 4 generators which are  $a_{1} $, $a_{2}$ and their adjoint $a_{1}^{\dagger} $, $a_{2}^{\dagger}$. Once again the operation  $\dagger$ maps $i\longrightarrow -i$ and $dx^{i}\longrightarrow -dx^{i}$.  Using the compactification on   $\mathbb{T}^4$, the  generators can be constructed  as follows
\begin{equation}
a_{1}=\frac{1}{2}(-dx^{1}+idx^{2}), \quad a_{2}=\frac{1}{2}(-dx^{3}+idx^{4})
\end{equation}
and \begin{equation}
a_{1}^{\dagger}=\frac{1}{2}(dx^{1}+idx^{2}), \quad a_{2}^{\dagger}=\frac{1}{2}(dx^{3}+idx^{4}).
\end{equation}
Similarly as  $Cl(2)$ and $Cl(4)$, the set of ladder  operators of $Cl(6)$  is $\lbrace a_{1},a_{2},a_{3},a_{1}^{\dagger},a_{2}^{\dagger}, a_{3}^{\dagger} \rbrace$ where
\begin{equation}
a_{1}=\frac{1}{2}(-dx^{5}+idx^{4}), \quad a_{2}=\frac{1}{2}(-dx^{3}+idx^{1}) \quad a_{3}=\frac{1}{2}(-dx^{6}+idx^{2})
\end{equation}
and
\begin{equation}
a_{1}^{\dagger}=\frac{1}{2}(dx^{5}+idx^{4}), \quad a_{2}^{\dagger}=\frac{1}{2}(dx^{3}+idx^{1}) \quad a_{3}^{\dagger}=\frac{1}{2}(dx^{6}+idx^{2})
\end{equation}
satisfying the above Clifford structures.
\subsection{$Cl(n)$  and dyonic black  objects}
The above  dyonic  black object  construction can be made  using D-branes moving on non trivial cycles.  In the associated compactification,  one can distinguish two
types  of cycles  namely  electric and magnetic cycles  noted  by  ${\cal C}^a$ and ${\cal D}^a$, respectively
\begin{equation}
\{[{\cal C}^\alpha ], \; \alpha=0,\ldots, 2^{n-1}-1 \},   \quad                \{[ {\cal D}^\alpha], \; \alpha=0,\ldots, 2^{n-1}-1\}.
\end{equation}
These cycles are dual to $\zeta$ and $ \eta$ forms on $\mathbb{T}^n$ defined by
\begin{equation}
\int_{{\cal C}^\alpha}\zeta^\beta=\delta^\alpha_\beta, \qquad  \int_{{\cal D}^\alpha}\eta^\beta=\delta^\alpha_\beta.
\end{equation}
 such that
 \begin{equation}
\int_{\mathbb{T}^n}\zeta^\alpha\wedge \eta^\beta=\delta^\beta_\alpha.
\end{equation}
The electric   vectors  determine
a basis for the  $2^{n-1}$  abelian  $(p+1)$-gauge
field $C_{p+1}$ obtained by integrating the ten dimensional forms on  ${\cal C}^\alpha$.  Similarly,  the magnetic    vectors  determine
a basis for the  $2^{n-1}$  abelian vector fields $(q+1)$-gauge
field $C_{q+1}$ obtained by integrating the ten dimensional forms on  ${\cal D}^\alpha$.   Under these  abelian gauge fields,  associated with $ \mbox{G}_{dyon}=  {\mbox{U}}(1)_e^{2^{n-1}}\times {\mbox{U}}(1)_m^{2^{n-1}} $  dyonic  gauge symmetry,
  the dyonic black objects in  generic  cohomology class   $[X]$, on $\mathbb{T}^n$,   are given by
\begin{equation}
 [X ]=  \sum_{\alpha=0}^{2^{n-1}-1}P_\alpha[{\cal C}^\alpha ]+ Q_\alpha [ {\cal D}^\alpha]
\end{equation}
 which carry  $P_\alpha$  electric and  $Q_\alpha$   magnetic charges. The
 dual of this  cycle is  a  general  element of  $Cl(n)$  with real coefficient associated with physical charges.  Motivated  by ideals of
 $Cl(n)$ used in  SM of particle physics \cite{26,27}, we associate to each  $ [X ]$    the following black object state
  \begin{equation}
 |\psi\rangle= Cl(n)|0\rangle
\end{equation}
where  $|0\rangle$ is considered  here as  a vacuum state. To be precise, the state $|\psi\rangle$  is  mapped to the class  $[X ]$
\begin{equation}
[X ]  \leftrightarrow  |\psi\rangle.
\end{equation}
In terms  of the  ladder  operators $a_\ell$, this state can take the following form
\begin{equation}
 |\psi\rangle= \sum_{e_1,\ldots,e_n=0,1} p_{e_1\ldots e_n}\prod_{\ell=1}^{\frac{n}{2}} a_\ell^{e_\ell}\vee {a^{\dagger}}_\ell^{e_{\ell+\frac{n}{2}}}|0\rangle,
\end{equation}
where $ p_{e_1\ldots e_n}$ are  real coefficients  describing   electric and magnetic charges of dyonic solutions.  It should be noted that the ordering problem  can be absorbed by negative charges carried by D-brane objects.   Moreover, the electric/magnetic charge duality can be  assured by
\begin{equation}
p_{e_1\ldots e_n} \leftrightarrow  p_{\overline{e_1}\ldots \overline{e_n}}
\end{equation}
originated from the Hodge duality of ${\mathbb{T}^n}$.  In this way, these charges can be  nicely organized as follows
\begin{equation}
\left(\begin{array}{c}
Q_\alpha\\
P_\alpha
\end{array}\right) \equiv  \left(\begin{array}{c}
p_{e_1\ldots e_n}\\
p_{\overline{e_1}\ldots \overline{e_n}}
\end{array}\right), \qquad \alpha=0,\ldots, 2^{n-1}-1.
\end{equation}
 In order to illustrate this  procedure in detail we   consider lower dimensional cases.  We believe that the general case can be dealt with without ambiguities.   For $n=2$  associated with the  $Cl(2)$ structure, the situation is  simple. The  charges of dyonic black objets can be organized as follows
\begin{equation}
p_{e_1e_2}\equiv (Q_\alpha, P_\alpha), \quad \alpha=0,1, \quad e_i=0,1.
\end{equation}
A quick  examination shows that one can use the following ordering  index
\begin{equation}
 Q_\alpha= p_{\alpha0} \qquad  P_\alpha= p_{1\alpha}, \quad \alpha=0,1.
\end{equation}
This involves   two electric charges and two magnetic charges corresponding to the following  graphic operator representation
\begin{equation}
\begin{tabular}{lll}
& 1 &  \\
$a \vee 1 $ &  & $ 1 \vee a^{\dagger}$ \\
& $ a \vee a^{\dagger}$ &
\end{tabular}.
\end{equation}
In this  graphic   representation, the  operators (1, $a \vee 1$ ) correspond to
the (D$p$,  D$p+1/\mathbb{S}^1_1$)  brane system  carrying
electric charges,  under the $\mbox{U}(1)_e^2$  electric gauge symmetry. However, the  dual operators  ($1 \vee a^{\dagger},
 a \vee a^{\dagger}$)   are associated with   the (D$q+1/\mathbb{S}^1_2$,
D$q+2/\mathbb{T}^2$)  brane configuration  having two magnetic
charges, under the $\mbox{U}(1)_m^2$  magnetic gauge symmetry

 For  $n=4$, the situation is quite different involving extra indices.  It is not obvious to find an elegant  index notation. However, we  will consider  a simple one inspired by the previous case.  The charges are indexed as
 \begin{equation}
p_{e_1e_2 e_3e_4}\equiv (Q_\alpha, P_\alpha), \quad \alpha=0,\ldots, 7.
\end{equation}
 In this way, they can be arranged as follows
 \begin{equation}\left(\begin{array}{c}
p_{0000} \;\; p_{1000} \;\;  p_{0100} \;\;  p_{0010} \;\; p_{0001} \;\; p_{1100} \;\; p_{1010} \;\; p_{1001}    \\
p_{1111} \;\; p_{0111} \;\;  p_{1011} \;\;  p_{1101} \;\; p_{1110} \;\; p_{0011} \;\; p_{0101} \;\; p_{0110}
\end{array}\right) \equiv \left(\begin{array}{c}
Q_0 \;\; Q_1 \;\;  Q_ 2\;\; Q_3 \;\; Q_4 \;\; Q_5 \;\; Q_6 \;\; Q_7   \\
P_0 \;\; P_1 \;\;  P_ 2 \;\; P_3 \;\; P_4 \;\; P_5 \;\; P_6 \;\; P_7
\end{array}\right)
\end{equation}
generating  6-dimensional  dyonic black objets obtained from  the compactification of  string theory on  $\mathbb{T}^4$.  They are charged under  gauge fields associated with the $ \mbox{G}_{dyon}= {\mbox{U}}^{8}(1)_e\times {\mbox{U}}^{8}(1)_m $  dyonic gauge symmetry.
\section{Conclusion and discussions}
 In this work,  we  have   investigated dyonic black objects in arbitrary dimensions. Inspired by   real Hodge diagrams of the   toroidal compactifcations on  ${\mathbb{T}^n}$, we have proposed a new  class  of doublets  of dyonic black solutions formed by two different D-branes given by
 $ \left(\begin{array}{c}
p\\
6-n-p
\end{array}\right)$ objects. In particular, we have
pointed out a correspondence between toroidal cycles  and the  black
solutions carrying electric and magnetic  charges, under  $ \mbox{G}_{dyon}=  {\mbox{U}}(1)_e^{2^{n-1}}\times {\mbox{U}}(1)_m^{2^{n-1}} $ dyonic  gauge symmetry. It has been suggested that this   symmetry could be  associated with electrically charged
magnetic monopole solutions  in stringy model buildings of the SM  extensions.
More precisely, we have considered in some details such dyonic black  classes obtained from lower dimensional toroidal compactification, and we have found
that they are linked to  $Cl(n)$ Clifford algebras using the  vee product.
It  has been observed that the analysis presented here might be extended to  a class of  local Calabi-Yau manifolds. These manifolds  have been extensively studied in the geometric engineering method of quantum field theory, obtained  from the compactification of higher dimensional  models including  string theory, M-theory, and F-theory \cite{29,30}. Some of them are known by canonical bundles on $n$-dimensional compact spaces given by   trivial  fibrations of $n$ copies  of the projective space  $\mathbb{CP}^1$ namely
\begin{equation}
{\cal M}^n=  \underset{n}{\underbrace{\mathbb{CP}^1 \times \ldots \times  \mathbb{CP}^1}}.
\end{equation}
The compactification of string theory on such manifolds produces dyonic black  objects  in $8-2n$-dimensional space-time.  Like toroidal compactifications, we are thus expecting that a similar analysis can take place by
replacing  the circle $\mathbb{S}^1$  by the one-dimensional complex  space $\mathbb{CP}^1$
\begin{equation}
\mathbb{S}^1 \rightarrow   \mathbb{CP}^1.
\end{equation}
Concretely, one should also expect to be able to  engineer   dyonic  black objects  in terms of  even dimensional D-branes producing models with less supersymmetric charges.\\
This work opens up for further discussions. We could cute  some of them.  First, it would  be interesting to investigate  geometries  based on  odd dimensional toroidal compactifications.  Second, it  should be nice to find a  link with fermionic coherent state theory shearing certain similarities with the dyonic black objects dealt with  in the present work. On the other hand,   one should also try   to establish a  relation or a link between graph theory and  such dyonic black objects  via Clifford algebra structures.

  Before closing  this discussion, we would like to make a remark  on  monopoles. It is recalled that the search   of such objects is the subject of many scientific  efforts  including
 the ATLAS-LHC experiment.  In  string theory compactification however, the  monopole/dyon sectors cannot
be separated and could be    studied in  realistic string-inspired SM extensions. In this  framework, one may think about a possible  doublet decomposition
\begin{equation}
\left(\begin{array}{c}
p\\
6-n-p
\end{array}\right)=\left(\begin{array}{c}
p\\
p
\end{array}\right)+\left(\begin{array}{c}
0\\
6-n-2p
\end{array}\right).
\end{equation}
In this  way, the last factor involves only magnetic charges which could be useful in such investigations from  stringy monopole analysis using D-branes wrapping non trivial magnetic   cycles.  We leave these investigations  for future works.

\end{document}